\documentclass[letterpaper,11pt]{article}

\usepackage{amsmath}
\usepackage{amssymb}
\usepackage{authblk}
\usepackage{booktabs}
\usepackage{cleveref}
\usepackage[margin=1in]{geometry}
\usepackage{nomencl}
\usepackage{siunitx}
\usepackage{subcaption}
\usepackage[disable]{todonotes}

\usepackage{lmodern}
\usepackage[T1]{fontenc}

\usepackage[
    backend=biber,
    citestyle=numeric-comp,
    language=english,
    sortcites=true,
    sorting=none,
    style=numeric-comp,
]{biblatex}
\makeatletter
\newlength{\bibsep}{\@listi \global\bibsep\itemsep \global\advance\bibsep by\parsep}
\makeatother
\newcommand{\phSc}[2] {^\mathrm{#1}_\mathrm{#2}}

\begin{document}

\title{Assessing Parameterized Geometric Models of Woven Composites using Image-Based Simulations}

\author[1,2]{Collin W. Foster}

\author[2]{Lincoln N. Collins}

\author[1]{Francesco Panerai\thanks{fpanerai@illinois.edu}}

\author[2]{Scott A. Roberts\thanks{Corresponding author, sarober@sandia.gov}}

\affil[1]{Department of Aerospace Engineering, University of Illinois at Urbana-Champaign, Urbana, IL 61801, USA}
\affil[2]{Engineering Sciences Center, Sandia National Laboratories, Albuquerque, NM 87185, USA}

\maketitle

\begin{abstract}
    Mesoscale simulations of woven composites using parameterized analytical geometries offer a way to connect constituent material properties and their geometric arrangement to effective composite properties and performance. However, the reality of as-manufactured materials often differs from the ideal, both in terms of tow geometry and manufacturing heterogeneity. As such, resultant composite properties may differ from analytical predictions and exhibit significant local variations within a material.  

    We employ mesoscale finite element method simulations to compare idealized analytical and as-manufactured woven composite materials and study the sensitivity of their effective properties to the mesoscale geometry. Three-dimensional geometries are reconstructed from X-ray computed tomography, image segmentation is performed using deep learning methods, and local fiber orientation is obtained using the structure tensor calculated from image scans. Suitable approximations to composite properties, using analytical unit cell calculations and effective media theory, are assessed. Our findings show that an analytical geometry and sub-unit cell geometry provide reasonable predictions for the effective thermal properties of a multi-layer production composite.
\end{abstract}

\section{Introduction}

The development of high temperature materials, such as those used to protect space vehicles subjected to the extreme aerothermal heating of atmospheric flight, is of growing research interest across a variety of communities. Fiber-based woven polymeric composites are often deployed to absorb and redirect the aerodynamic heating experienced during atmospheric flight through decomposition and ablation \cite{Weng2015}, while maintaining the mechanical strength necessary to carry harsh mechanical loads. The effective thermophysical properties of these composites are dictated by their microscale (fiber) and mesoscale (tow and matrix) composition and geometry, and determine their performance as a thermal protection system \cite{Shim2002, Tian2012}. 

Construction of analytical composite weaves enables convenient, rapid design studies and optimization of woven composites. TexGen \cite{Long2011}, WiseTex \cite{Verpoest2005}, or sinusoidal \cite{Naik1994, Collins2021} generation of weave geometries present configurable options to evaluate the impact of mesoscale composite geometry on properties and performance. Although useful tools for idealized or dry weaves, the generation of symmetric cross-sectional geometry often misrepresents as-manufactured cured weaves. Additionally, using non-nested fabric tows can result in incorrect estimation of the weave and fiber volume fractions leading to incorrect simulation results and effective properties when compared to the real material \cite{Sevenois2016}. 

Alternatively, tomographic imaging of composites enables direct numerical predictions of as-manufactured composite properties, including uncertainty estimates \cite{Krygier2021}. Performing simulations directly on composite geometries allows for obtaining valuable structure-property relations and performance estimates without expensive testing and characterization, greatly reducing the time required for design and optimization. High resolution X-ray computed tomography (XCT)\nomenclature{XCT}{X-ray Computed Tomography} offers a non-destructive sub-micron resolution capability that is used to evaluate complicated 3D woven composites of various compositions at multiple length scales \cite{Bale2012, Maire2014}. Regions of interest are resolved with XCT that can then be matched with relevant experimental property data to infer predictions on scaled systems of similar materials \cite{Panerai2017, Roberts2014, Ferguson2018, Trembacki2018, Mazars2017, Saucedo-Mora2017, Semeraro2021, Semeraro2020, Straumit2016}. 

However, XCT has limitations and difficulties ranging from data collection and instrumentation to the subsequent image analysis required for image-based simulation \cite{Haboub2014, Barnard2018}. Woven composites are 3D anisotropic composites that can have difficult to replicate weave architectures that call for high-resolution XCT for proper multi-scale characterization \cite{Venkatapathy2011, Beverly2019}. This is an issue, as often a compromise must be made between field of view (FOV)\nomenclature{FOV}{Field of View} and voxel resolution for bench-top or lab-standard XCT devices, often limiting high-resolution studies to sub-unit cell geometries. Segmenting and meshing tomographic data of woven composites for use in finite element simulations becomes more difficult with weave structure complexity. Composites are often cured in vacuum sealed or hot-pressed environments, resulting in large compaction of the fabric weave, nesting between layers, and local displacements and overlap within the weave, all contributing to difficulties in mesoscale geometric reconstruction.

This study examines how well a XCT image of a single-layer (SL)\nomenclature{SL}{Single Layer} woven composite is represented by an analytical geometry over relevant thermal and topological properties. Initially, we investigate whether a SL woven composite can approximate a larger, multi-layer (ML)\nomenclature{ML}{Multi-layer} coupon specimen of similar weave architecture in image-based simulations. The relevant measurements for defining the analytical geometry are then harvested directly from the as-manufactured SL. In evaluating how well analytic geometries can represent the image-based ones, we also probe the most influential geometric characteristics considered in weave generation through a Global Sensitivity Analysis (GSA)\nomenclature{GSA}{Global Sensitivity Analysis} via Latin Hypercube Sampling (LHS)\nomenclature{LHS}{Latin Hypercube Sampling}. We present results of these comparisons using composite density, fiber volume fraction, specific heat, and effective thermal conductivity as quantities of interest (QOI)\nomenclature{QOI}{Quantities of Interest}.

\section{Material and methods}

\subsection{Experimental}
    The material examined in all datasets is a compression-molded silica phenolic composite.  It is composed of an 8-harness satin (8HS)\nomenclature{8HS}{8-Harness Satin} silica fiber cloth (Refrasil, Hitco Carbon Composites, California USA) impregnated with a phenolic resin (Durite SC-1008, Hexion Specialty Chemicals, Inc., Kentucky USA) and pressed/cured according to Durite manufacturing specifications. The first sample introduced is a single-layer, at 0.46 mm total thickness. The second geometry is a multi-layer\nomenclature{ML}{multi-layer} coupon, with an average layer thickness of 0.42 mm and a total thickness of 5.3 mm following the same manufacturing procedure as the SL geometry.

    The materials were scanned with a Bruker SkyScan 1272 at pixel sizes of \SI{2.0}{\micro\meter} for the SL and \SI{1.8}{\micro\meter} for the ML coupon, enabling clear resolution of the mesoscale. The overall dimensions of the obtained images were \qtyproduct{5.2 x 6.4 x 0.46}{\milli\meter} and \qtyproduct{3.7 x 3.5 x 5.3}{\milli\meter} for the SL and ML coupon, respectively. The resulting resolution, image texture, and phase contrast are sufficient for a variety of segmentation and image analysis techniques. An example volumetric reconstruction of the raw dataset is presented in \cref{fig:pipeline:recon} and the remaining binarized domains are found later in \cref{fig:geometries}. We emphasize that the chosen material specimens and associated XCT scans are very far from ideal, with irregularities, imperfections, and limitations representative of as-manufactured parts. 

\subsection{Computational Methods}

\begin{figure*}
    \centering
    \includegraphics[width=\linewidth]{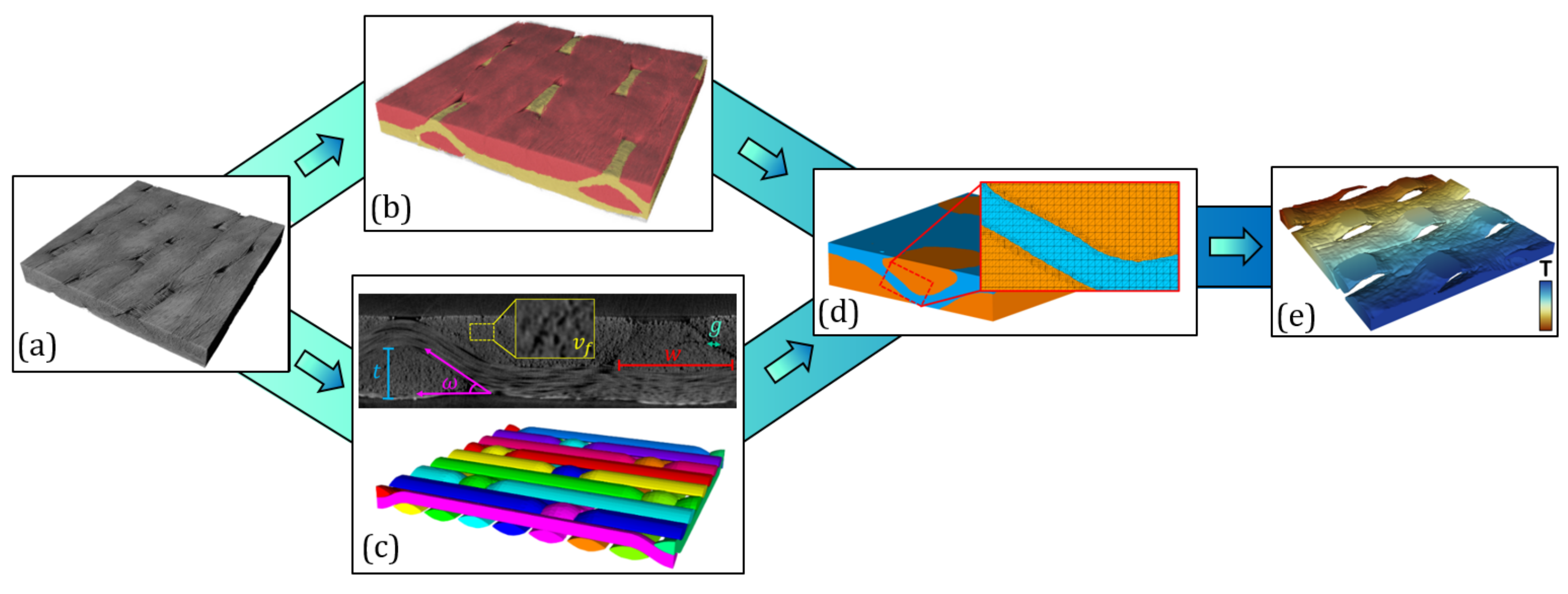}
    {
        \phantomsubcaption\label{fig:pipeline:recon}
        \phantomsubcaption\label{fig:pipeline:seg}
        \phantomsubcaption\label{fig:pipeline:analytical}
        \phantomsubcaption\label{fig:pipeline:mesh}
        \phantomsubcaption\label{fig:pipeline:sim}
    }
    \caption[Workflow for analytical and image-based woven composite simulations illustrated for a SL composite.]{Workflow for analytical and image-based woven composite simulations illustrated for a SL composite.  (a) 3D reconstruction of the raw XCT image.  (b) Three-phase image segmentation.  (c) Analytical geometry informed by parameters extracted from the XCT image: tow thickness ($t$), waviness ($\omega$), fiber packing ratio ($v_{f}$), tow width ($w$), and tow gap ($g$). (d) Volume mesh.  (e) Physics simulation result.  For image-based simulation the workflow is a-b-d-e. For the analytical model, the workflow is a-c-d-e.}
    \label{fig:pipeline}
\end{figure*}
  
    Our developed simulation workflow is summarized in \cref{fig:pipeline}. The reconstruction of the SL image geometry is captured as the raw XCT dataset without segmentation (\cref{fig:pipeline:recon}). The XCT data gives the clear resolution of the grayscale values needed for texture based machine-learning segmentation of the warp, weft, and matrix classes. We utilize a U-Net \cite{Ronneberger2015} deep convolutional neural network implemented in Dragonfly \cite{Dragonfly2020} to segment the warp and weft tows from the matrix (\cref{fig:pipeline:seg}). A 3D median filter is applied to fill voids and isolate the tows as the primary mesoscale features. Additionally, a Gaussian blur is implemented to smooth the binarized image for more effective surface meshing.
    
\begin{figure*}
        \centering
        \includegraphics[width=\linewidth]{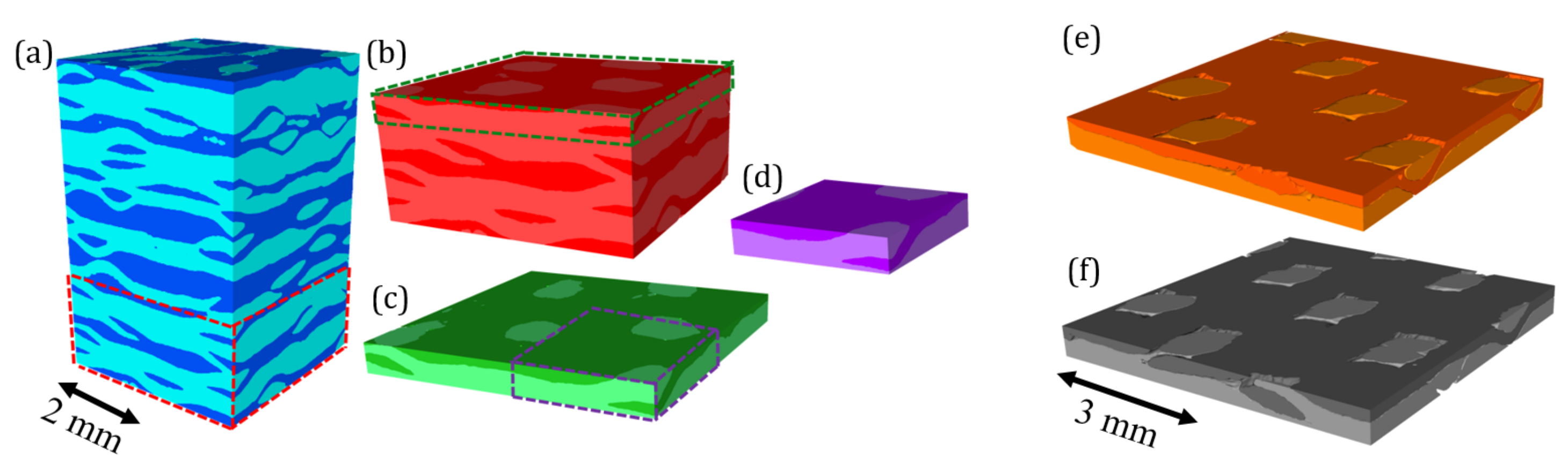}
        {
            \phantomsubcaption\label{fig:geometries:full}
            \phantomsubcaption\label{fig:geometries:half}
            \phantomsubcaption\label{fig:geometries:layer}
            \phantomsubcaption\label{fig:geometries:sub}
            \phantomsubcaption\label{fig:geometries:weft}
            \phantomsubcaption\label{fig:geometries:warp}
        }
        \caption[Finite element meshes from image-based geometries.]{Finite element meshes from image-based geometries. Warp and weft direction separated by color saturation. (a) - (d) represent the full, half, SL, and sub unit-cell coupon geometries with bounding boxes depicting subdomain selection for the entire geometry. (e) and (f) are the weft- and warp-aligned SL meshes. In all figures, the resulting matrix domain is excluded.}
        \label{fig:geometries}
    \end{figure*}

    Each dataset is surface meshed with a Lewiner Marching Cubes algorithm available through the Python library scikit-image to produce a Standard Tessellation Language (STL)\nomenclature{STL}{Standard Tessellation Language} file \cite{VanDerWalt2014}. The resulting STLs then are used to create volumetric meshes using the Conformal Decomposition Finite Element Method (CDFEM) (\cref{fig:pipeline:mesh}) \cite{Noble2010, Roberts2018}. A rectangular domain of prescribed dimensions is discretized into tetrahedral elements for calculating a signed level-set distance function, $\phi$, where $\phi=0$ represents the location of the STL surface. New nodes are added on edges of the elements intersecting this interface where $\phi=0$. Elements containing these new nodes are subsequently decomposed into child elements that conform to the interface. The resulting finite element meshed geometries used for this study are the SL image geometry shown in \cref{fig:geometries:warp} \& \cref{fig:geometries:weft} and ML coupon image geometry shown in \cref{fig:geometries:full}. 

    To capture the strong anisotropy inherent in woven composites, the local tow orientation vector (TOV)\nomenclature{TOV}{Tow Orientation Vector}, describing the longitudinal direction of filaments in the fabric, is determined via the image structure tensor. The local gradient of the grayscale intensity values of the unsegmented image is computed and the tow orientation corresponds to the smallest eigenvector \cite{Krause2010}. Erroneous voxel TOVs arising from the eigenvector calculation and image noise are blurred and aligned within each segmented domain such that the orientation is consistent throughout each warp/weft classification using the masks generated from the U-Net segmentation. However, the compact nature of the weaves results in orientation errors near the warp/weft boundaries from the image-gradient calculation over a neighborhood of each pixel. A Gaussian blur with standard deviation of 0.005 for the coupon and 0.3 for the SL of the resulting vector field minimized this error for the ML coupon and SL, respectively.

    Effective thermophysical properties of the composite are calculated through finite element simulations on the unit-cell geometries using the Sandia-developed SIERRA/Aria finite element code \cite{Sierra2020}. Nominal resin and fiber properties are chosen based on commercially-available products and are tabulated in \cref{tab:image}. Additionally, effective mesoscale axial and transverse conductivities for the tow regions are derived from effective media theory \cite{Chamis1983, Collins2021}. These values assume a unidirectional array of filaments with a prescribed packing density, informed by high resolution image-based measurements discussed in the following section. In simulations including TOV, the resulting tensor conductivity is rotated to align with the filaments. Without TOV, the axial result is assigned to the primary warp (or weft) direction, with transverse values applied in directions orthogonal to the axial direction. For effective density, specific heat, and fiber volume fraction of the composite, the volume of warp and weft regions are calculated using the finite element meshes, allowing for averaging of constituent material properties.

    \begin{table}
        \centering
        \caption{Nominal mesoscale properties for image-based simulations.}
        \label{tab:image}
        \begin{tabular}{@{}llSl@{}}
            \toprule
            Property & Units & {Value} & Ref. \\
            \midrule
            Fiber Axial Conductivity & \si{W/(m.K)} & 1.4 & \cite{Combis2012} \\
            Fiber Transverse Conductivity & \si{W/(m.K)} & 1.4 & \cite{Combis2012} \\
            Fiber Packing Ratio & & 0.674 & \cref{fig:lhs_dist:packing} \\
            Resin Conductivity & \si{W/(m.K)} & 0.215 & \cite{Mottram1987ii, Clayton1969} \\
            Tow Axial Conductivity & \si{W/(m.K)} & 1.013 & \cite{Chamis1983} \\ 
            Tow Transverse Conductivity & \si{W/(m.K)} & 0.616 & \cite{Chamis1983} \\               
            \bottomrule
        \end{tabular}
    \end{table}

    Thermal conductivity is calculated through separate simulations for the in the in-plane (IP)\nomenclature{IP}{In-plane} and out-of-plane (OOP)\nomenclature{OOP}{Out-of-plane} directions assuming steady-state Fourier's heat conduction \cite{Collins2021} and solving
    \begin{align}
        \nabla\cdot\left(\mathbf{k}' \cdot \nabla T\right) = 0\text{    in   }\Omega',
        \label{eq:fourier}
    \end{align}
    where $\mathbf{k}'=\mathbf{k}\phSc{t}{},\mathbf{k}\phSc{m}{}$ denotes the thermal conductivity tensor of the tow ($\Omega\phSc{t}{}$) and matrix ($\Omega\phSc{m}{}$) phases in the unit cell geometry and $T$ is the temperature. In each direction, a thermal gradient across the domain is applied through Dirichlet boundary conditions on opposing faces, with the remaining surfaces being adiabatic. Symmetry in the resulting temperature field is established with the symmetric unit-cells. The resulting heat flux, $Q$, through the the low-temperature boundary, $\partial_o\Omega$ with outward normal $\hat{n}$, is generated using the temperature solution: 
    \begin{align}
        Q = \frac{-\int_{\partial_o\Omega} \left(\mathbf{k}' \cdot \nabla T\right)\cdot\hat{n} \,\mathrm{dA}}{\int_{\partial_o\Omega}\,\mathrm{dA}}.
        \label{eq:flux}
    \end{align}
    This allows for the calculation of effective thermal conductivity, $k^*$ \cite{Hashin1979}:
    \nomenclature{$k^*$}{effective thermal conductivity}
    \begin{align}
        \label{eq:eff_cond}
        k^* = -\frac{L}{\Delta T}Q,
    \end{align}
    where $L$
    \nomenclature{$L$}{domain dimension}
    denotes the simulation domain dimension parallel to the temperature gradient, and $\Delta T$ is the difference between the enforced temperature boundary conditions.  

    The governing equation, \cref{eq:fourier}, is solved using a Galerkin finite-element scheme with linear basis functions. Newton's method is implemented for the nonlinear iterative solution, and the resulting linear system is solved using the generalized minimal residual (GMRES) method and an incomplete LU preconditioner from the Trilinos library \cite{trilinos}.

    Concurrently, visualized in \cref{fig:pipeline:analytical}, necessary geometric features for generating analytic geometries are manually extracted from the unprocessed warp- and weft-aligned SL images. These measurements establish parameter distributions that are then probed through the analytical study. Given a selection of geometric parameters, a system of sinusoidal piecewise functions is used to directly generate a surface description of an entire 8HS weave \cite{Collins2021}. The functional form of the description is adapted from \cite{Naik1994}. The weave is characterized by tow width $w$ and thickness $t$ assuming a balanced weave. The cross section and path shape are determined by length $L_u=u\cdot w$, dependent on the undulation of a tow $u$ and by the gap $g$ expressed by percentage of tow width. The tow cross-section and path through the unit cell take the form
    \begin{align}
    \label{eq:cross_sec_geo}
    f_\text{cross}(x) = \left\{
    \begin{array}{lcrcl}
    \pm \frac{t}{2}\sin\left(\frac{\pi}{L_u}x\right) & : & 0 \leq & x & \leq \frac{L_u}{2}\\
    \pm \frac{t}{2} & : & \frac{L_u}{2} <  & x & \leq \left(w - \frac{L_u}{2}\right) \\
    \mp \frac{t}{2}\sin\left(\frac{\pi}{L_u}(x-w)\right) & : & \left(w - \frac{L_u}{2}\right) < & x & \leq w\\
    \end{array}
    \right.
    \end{align}
    and
    \begin{align}
    \label{eq:path_geo}
    f_\text{path}(z \pm 2a) = \left\{
    \begin{array}{lcrcl}
    \frac{t}{2}\sin\left(\frac{\pi}{L_u}z\right) & : & 0 \leq & z & \leq \frac{L_u}{2}\\
    \frac{t}{2} & : & \frac{L_u}{2} < & z & \leq a - \frac{L_u}{2} \\
    -\frac{t}{2}\sin\left(\frac{\pi}{L_u}(z-a)\right) & : & a - \frac{L_u}{2} < & z & \leq a + \frac{L_u}{2}\\
    -\frac{t}{2} & : & a + \frac{L_u}{2} < & z & \leq 2a - \frac{L_u}{2}\\
    \frac{t}{2}\sin\left(\frac{\pi}{L_u}(z-2a)\right) & : & 2a - \frac{L_u}{2} & z & \leq 2a
    \end{array}
    \right..
    \end{align}
    Here, $a$ represents half of the unit cell width: $a=w(1+g)$. A complete tow surface is constructed by summing the centerline and cross-section functions for $(x,z)$:
    \begin{align}
    y = f_\text{tow}(x,z) = f_\text{path}(z) \pm f_\text{cross}(x), (x,z) \in (0,w)\times(0,2a).
    \end{align}
    Waviness $\omega$ describes the average slope of the undulating portion of the yarn in terms of the geometric parameters and will be used for analysis of the composite properties:
    \begin{equation}
    \label{eq:waviness}
    \omega = \frac{2t}{wu}.
    \end{equation}
    These geometries follow the same workflow as the image-based datasets, resulting in effective thermal conductivity simulations for the IP and OOP directions.

    The geometric properties were directly sampled from the SL image geometry to inform the sensitivity analysis performed in Dakota \cite{Dakota2020}. The parameters required for the simulated geometry are the fiber packing ratio and the tow thickness, gap, width, and undulation. Undulation is calculated from the waviness \cref{eq:waviness}, using the width, height, and sloped portions of the tow. Fiber packing ratio is approximated using micro-scale resolution cross sections of the tows, and thresholding out the fiber volume fraction. Shearing of as-manufactured 8HS composites lead to a non-perpendicular layup and slight biasing in the fabric causing differences between the warp and weft tows.
    
    Much like the image-based simulations, this study calculates the following QOIs through simulations on analytical weave geometries: fiber volume fraction, specific surface area, specific contact area, density, specific heat, IP and OOP conductivity. We capture local geometry and property variation seen in the image-based geometries by generating and performing simulations on analytical 8HS unit cell geometries parameterized by measurements of the images through LHS \cite{Dakota2020} for 1500 realizations. A polynomial chaos expansion (PCE)\nomenclature{PCE}{Polynomial Chaos Expansion} surrogate model \cite{Sudre2008} is then developed to describe the resulting property distributions and allow for a sensitivity analysis relating the material parameter space to composite effective properties through Sobol’ indices.

\section{Results and discussion}

\subsection{Image-based simulation}
 
    An evaluation of decreasing domain size of a compact ML composite and its effect on effective thermal conductivity is performed and compared to a similar SL sample. The image-based geometries are represented in their two-tone warp-weft segmentation configurations in \cref{fig:geometries}. The complete coupon image geometry presents an opportunity to study the necessity of including multiple layers in effective property simulations, as opposed to one or even a sub-unit cell geometry. As such, the full dataset is cropped to include all layers (\cref{fig:geometries:full}), approximately four layers (\cref{fig:geometries:half}), approximately one layer (\cref{fig:geometries:layer}), and a quarter layer (\cref{fig:geometries:sub}) for simulation. Performing a domain size study down to the sub-unit cell geometry provides insight to the utility of XCT scans with a large FOV at the compromise of voxel size. The high degree of weave compaction should be noted in each of the coupon geometries, with weave volume constituting nearly 99\% of the total domain volume after segmentation. Comparisons with reported fiber volume fractions can be estimated assuming a nominal fiber packing ratio within the tows, expressed in \cref{tab:image}.

    The remaining samples are the warp- and weft-aligned SL image geometries shown in \cref{fig:geometries:warp,fig:geometries:weft}. The thermal conductivity phase assignment assumes a balanced fabric with the weave directions aligned with the IP coordinate axes. However, the SL composite sample experienced a slight IP shear during fabrication, prompting an investigation of the degree of resultant anisotropy. The warp and weft components are measured to be off-orthogonal by approximately 10 degrees. For this reason, two separate near-unit cells: the warp- and weft-aligned orientations, are extracted where each fabric component aligns with a coordinate axis for effective-property calculations. Similar to the coupon image geometries, but without nesting between layers, the SL image geometries have a slightly lower weave volume fraction of 95\% with small visible gaps where the intertwining tows overlay.

    \begin{figure}
        \centering
        \includegraphics[width=\linewidth]{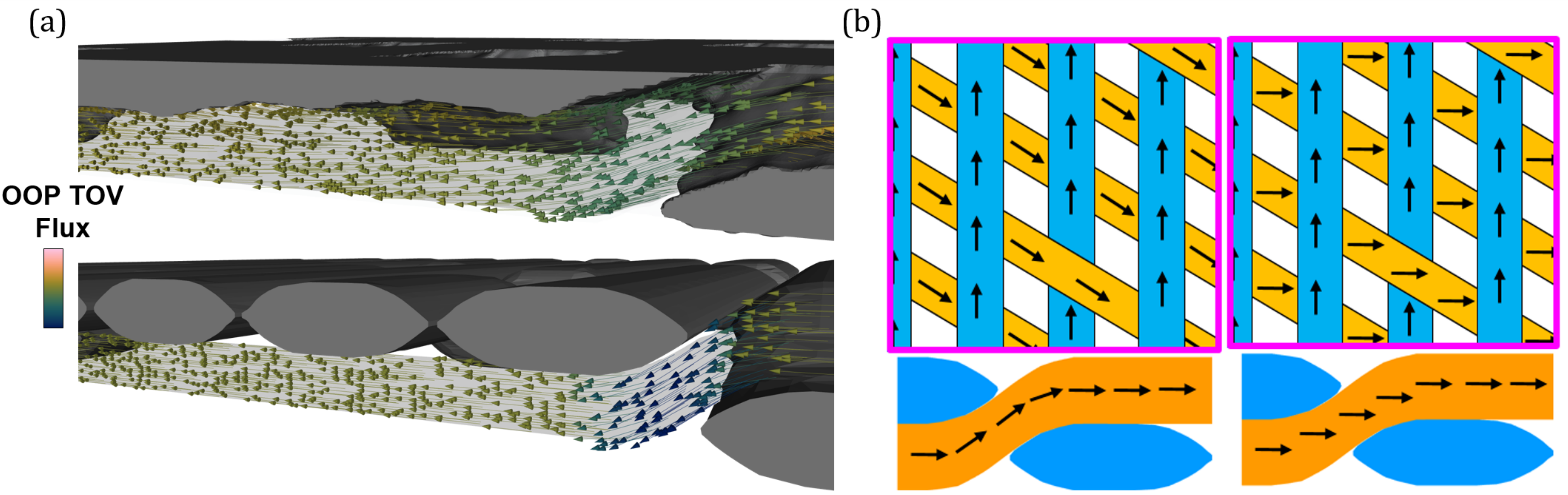}
        {
            \phantomsubcaption\label{fig:tov:model}
            \phantomsubcaption\label{fig:tov:example}
        }
        \caption[Local fiber orientation depicted by the tow orientation vectors (TOVs).]{Local fiber orientation depicted by the tow orientation vectors (TOVs) for the (a-top) image-based geometry and (a-bottom) analytical geometry using the structure tensor, and (b) simplified geometry (b-left) with TOVS and (b-right) without TOVs.}
        \label{fig:tov}
    \end{figure}

    The local material orientation is computed as a post-processing step after segmentation. Each voxel pertaining to a tow must have an assigned TOV to adequately represent the path of thermal transfer through a local conductivity tensor. \cref{fig:tov:model} shows the SL geometry for both the image-based and analytical results. The gray weft tows have a highly compact nature, making each indistinguishable from its neighbor in the image-based geometry above. The TOVs are colored by their OOP-component, qualitatively matching the tow path through the undulating portions and approaching zero in the flat portions of the weave. Implementing TOVs is an essential step in evaluating the thermal properties in woven composites with distinct OOP travel and IP shearing \cite{Semeraro2021}. 

    A simplified example of the effect of TOVs is shown in \cref{fig:tov:example}, where a weave without supplied TOVs is shown to the right with coordinate-aligned assignment of thermal conductivities. The left figure shows the application of TOVs, representing the shear layup with more truthful tensor orientations, resulting in a decrease in the weft and increase in the warp direction conductivity. The tow orientation vectors give a more accurate representation of undulation on the thermal conductivity tensor throughout the tows' path. Intuitively, this inclusion will increase conductivity in the OOP-direction as the undulating tow path directs more fibers away from the IP-directions, especially considering the tows' higher axial conductivity compared to the transverse direction \cite{Chamis1983}. The introduction of TOVs for nearly all cases analyzed reduced the spread in IP conductivity values. This is likely a result of the shearing of the fabric weave during layup and the TOVs accounting for the tows being shifted off the major axes. The inclusion of TOVs decreased IP conductivity in the weft-direction and increased it in the warp-direction for all cases.

    \begin{figure*}
        \centering
        \includegraphics[width=\linewidth]{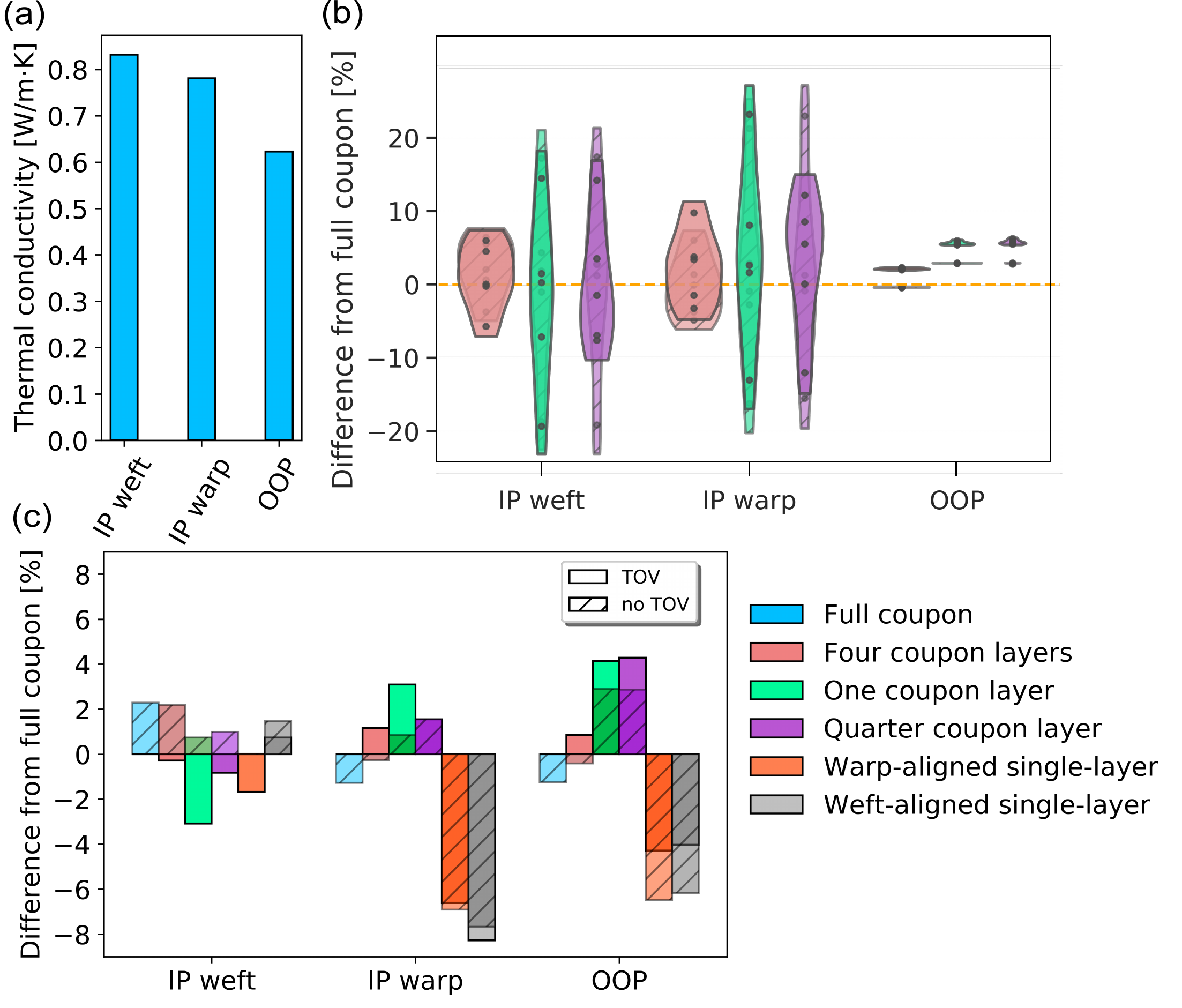}
        {
            \phantomsubcaption\label{fig:all_image_conductivities:coupon}
            \phantomsubcaption\label{fig:all_image_conductivities:error}
            \phantomsubcaption\label{fig:all_image_conductivities:comparisons}
        }
        \caption[Effective thermal conductivities for analyzed image-based geometries.]{Effective thermal conductivities for analyzed image-based geometries. (a) The calculated thermal conductivity in all three directions for the entire coupon geometry. (b) Variation in thermal conductivity when subdomains are sampled from different locations in full ML coupon. (c) Thermal conductivity comparison of selected coupon subdomains and the SL geometries relative to the value for the full ML coupon with TOV.}
        \label{fig:all_image_conductivities}
    \end{figure*}

    The results of the thermal conductivity simulations on the image-based geometries are presented in \cref{fig:all_image_conductivities}. Results are shown in reference to the full ML coupon with TOVs, which is seen as the most comprehensive/correct result because of its inclusion of all layers available (\cref{fig:all_image_conductivities:coupon}) and the highest accuracy material anisotropy because of the TOVs. 
    
    The ML coupon domain was sampled at five different locations for the four, one, and quarter layer subdomains to explore the spatial variability in thermal conductivity. Shown in \cref{fig:all_image_conductivities:error}, the larger domain captured in the four layers has less deviation compared to the one and quarter layer samples in the IP direction, indicating variation in fabric layup layer-to-layer. Although the difference in mean IP value of the one and quarter layer geometries compared to the full coupon was less than 4\%, it is important to note the possibility of potential outliers that are shown from the respective extrema. Simulating only a single smaller subdomain could introduce errors larger than 20\%. 
    
    The inclusion of TOVs in one coupon layer caused more deviation from the full in all directions, suggesting that the effects of IP fabric shearing and OOP undulation were accentuated in the smaller domains. Given the compact nature of all samples in this study, there was little OOP variation observed, although the one and quarter layers were less accurate overall, shown by the mean values in \cref{fig:all_image_conductivities:comparisons}. As a result, the observed deviation provides helpful insight when considering sampling subdomains when a representative volume element (RVE)\nomenclature{RVE}{Representative Volume Element} or full unit cell is not available. 

    The SL geometry and coupon subsets reveal that there are limitations of the structure tensor on very compacted composite samples. The SL geometries are closer to representing a complete 8HS RVE, and include a higher proportion of undulating tows that contribute to the difference in OOP conductivity when TOVs are introduced. The degree of anisotropy as a result of the manufacturing shear is small in the SL geometries, as indicated by the slight differences in IP conductivities when TOVs are introduced in \cref{fig:all_image_conductivities:comparisons}. 

    SL image geometries can approximate a coupon within 8\% (\cref{fig:all_image_conductivities:comparisons}), although small differences in effective thermal conductivity will occur if the coupon is not at least a unit cell representation in the IP directions. Consideration of potential outliers should be given when sampling one (or sub-unit cell) layer geometries, although when sampled and averaged the highest difference is seen in the OOP direction at 4\%. Larger domain inclusion (four layers) will likely result in more accurate property distributions. 

\subsection{SL analytical study}

    \begin{figure}
        \centering
        \begin{subfigure}{0.31\textwidth}
            \centering
            \includegraphics[width=\textwidth]{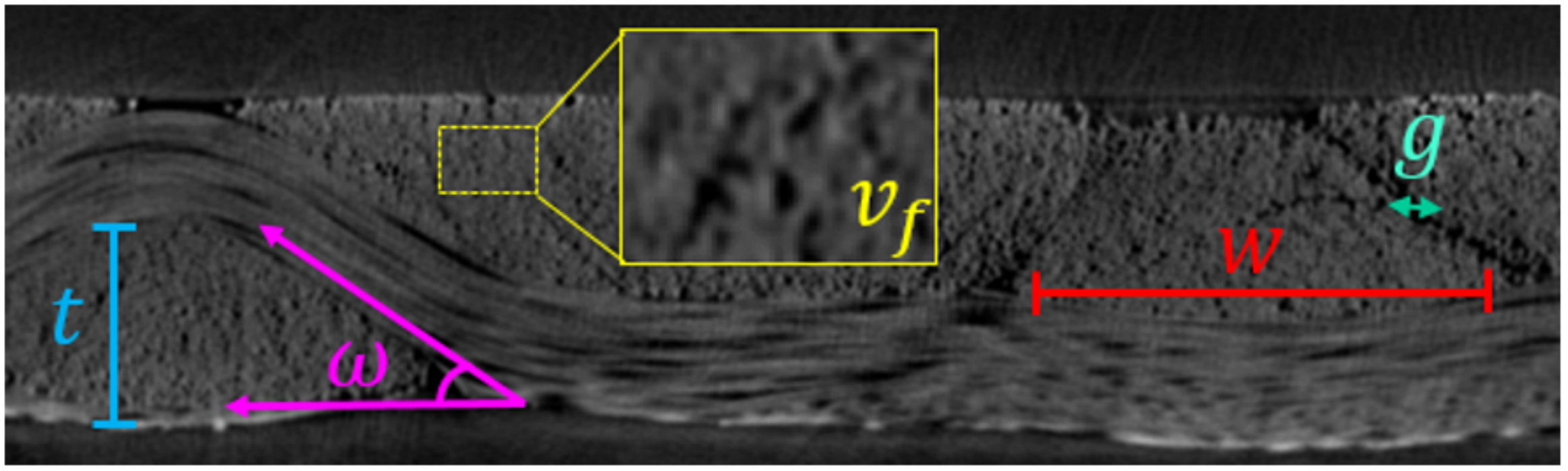}
            \caption{XCT reconstruction}
            \label{fig:lhs_dist:img}
        \end{subfigure}
        \begin{subfigure}{0.31\textwidth}
            \centering
            \includegraphics[width=\textwidth]{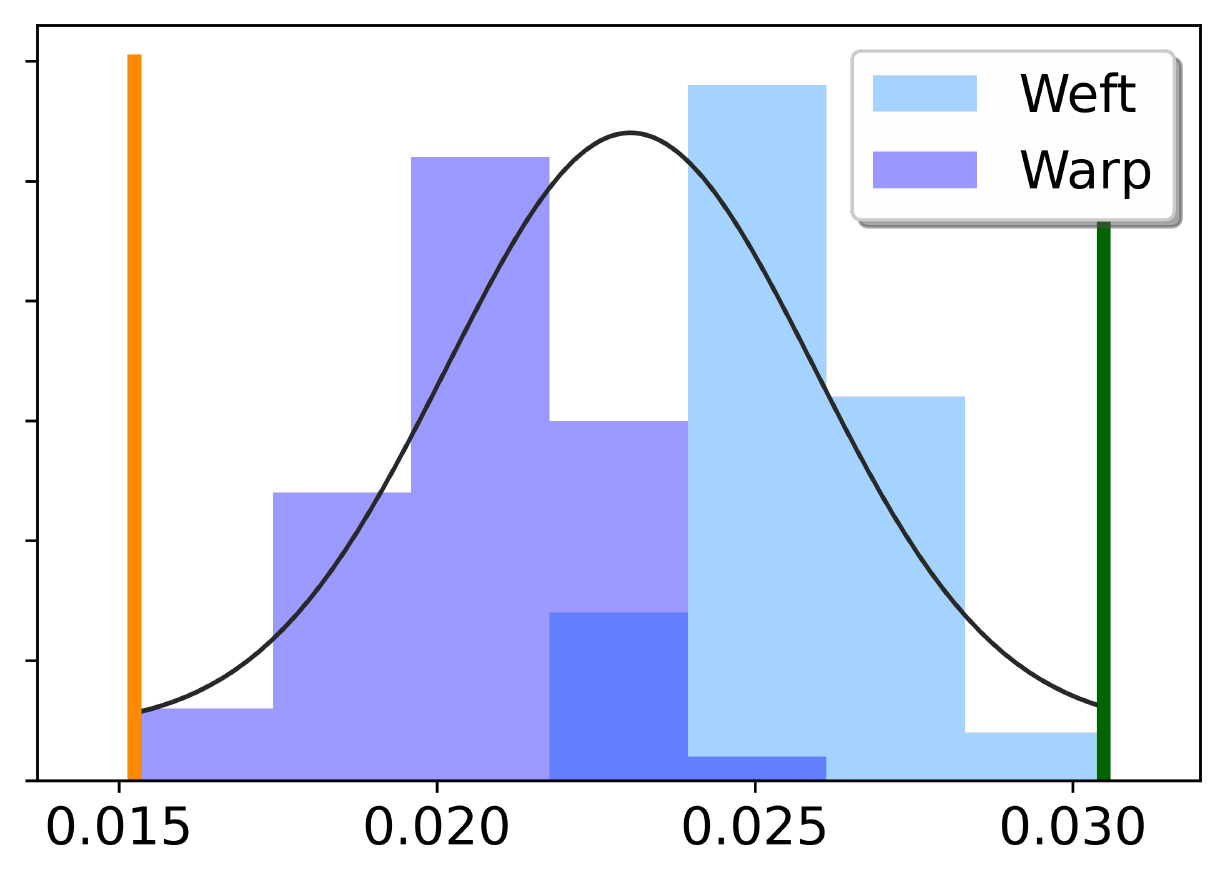}
            \caption{Tow thickness [\unit{\centi\meter}]}
            \label{fig:lhs_dist:t}
        \end{subfigure} 
        \begin{subfigure}{0.31\textwidth}
            \centering
            \includegraphics[width=\textwidth]{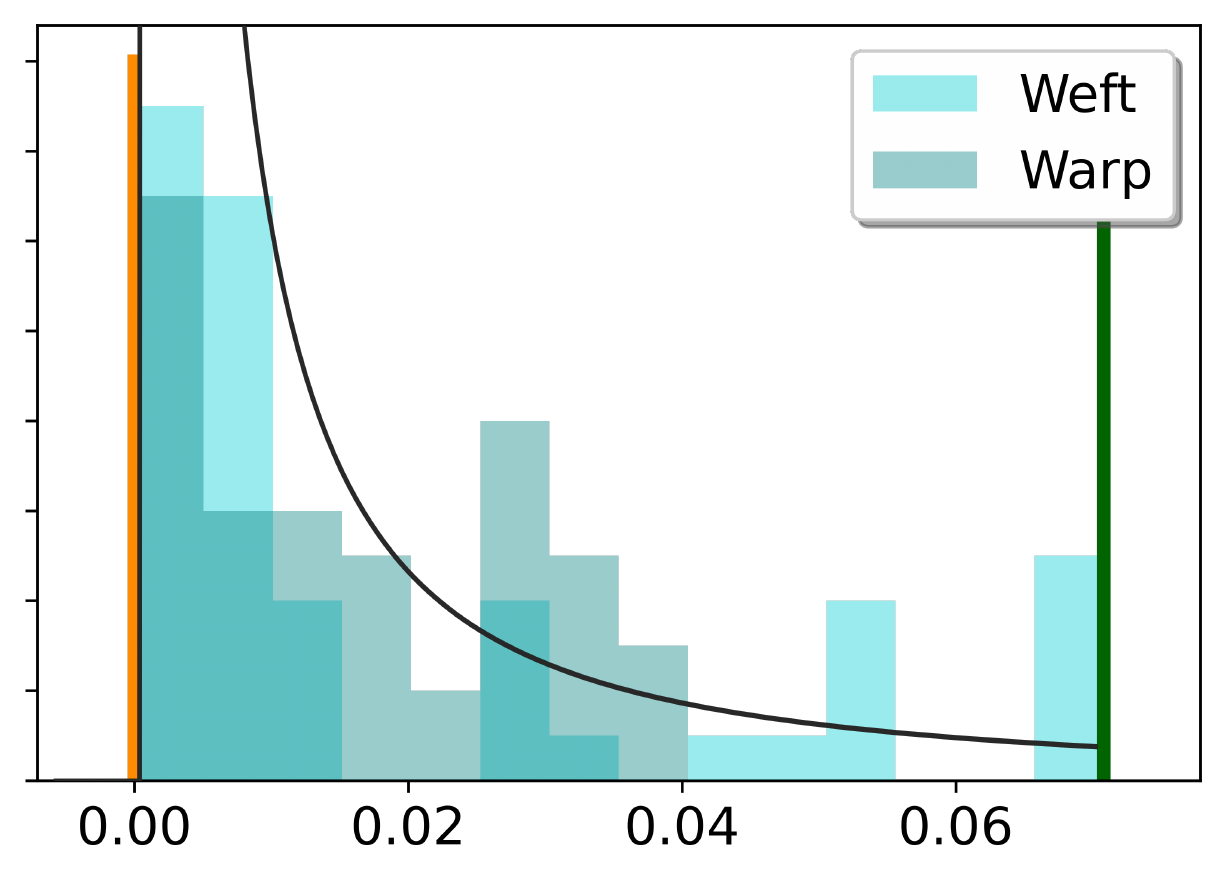}
            \caption{Tow gap [\unit{\centi\meter}]}
            \label{fig:lhs_dist:gap}
        \end{subfigure} \\
        \begin{subfigure}{0.31\textwidth}
            \centering
            \includegraphics[width=\textwidth]{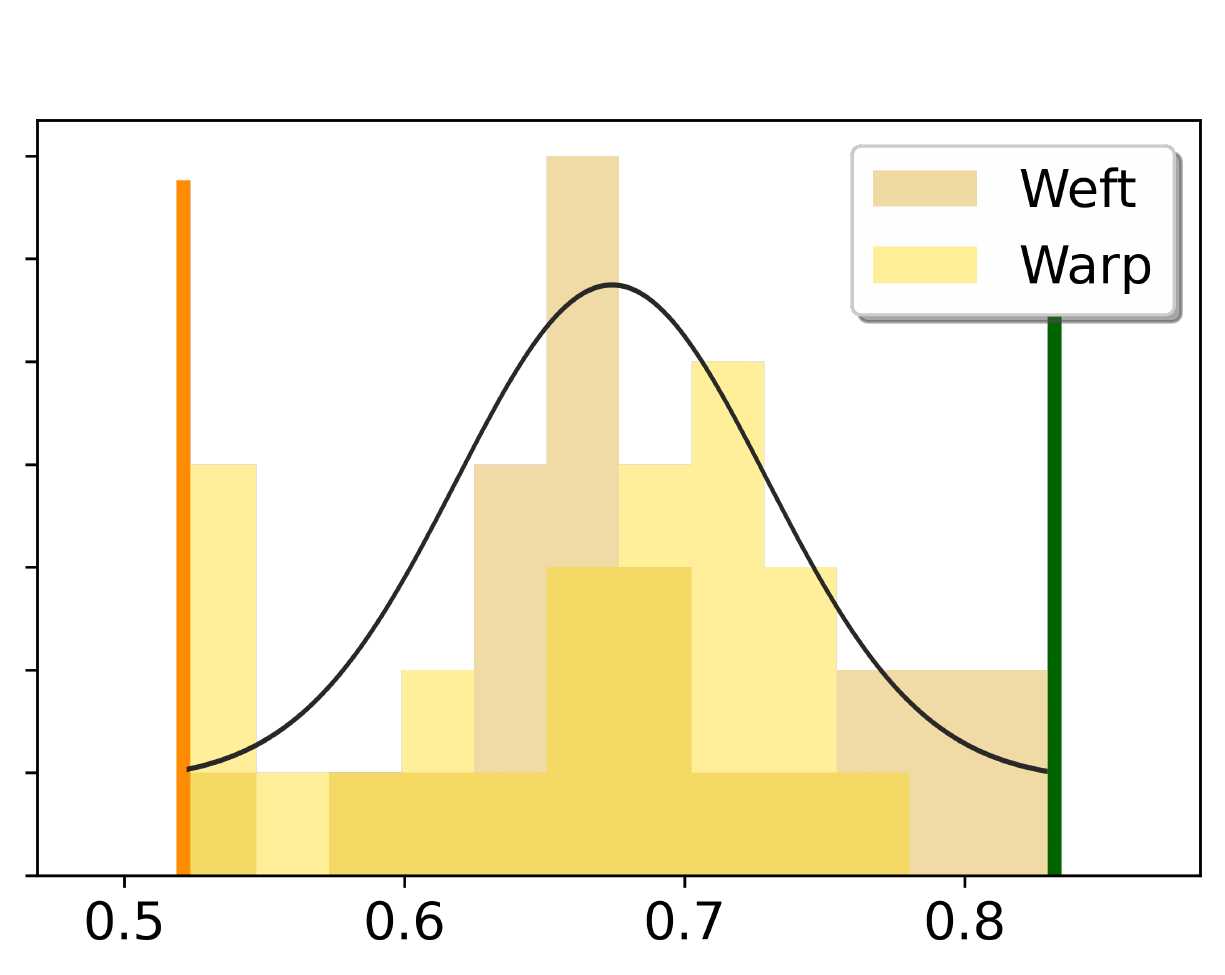}
            \caption{Fiber packing ratio}
            \label{fig:lhs_dist:packing}
        \end{subfigure}
        \begin{subfigure}{0.31\textwidth}
            \centering
            \includegraphics[width=\textwidth]{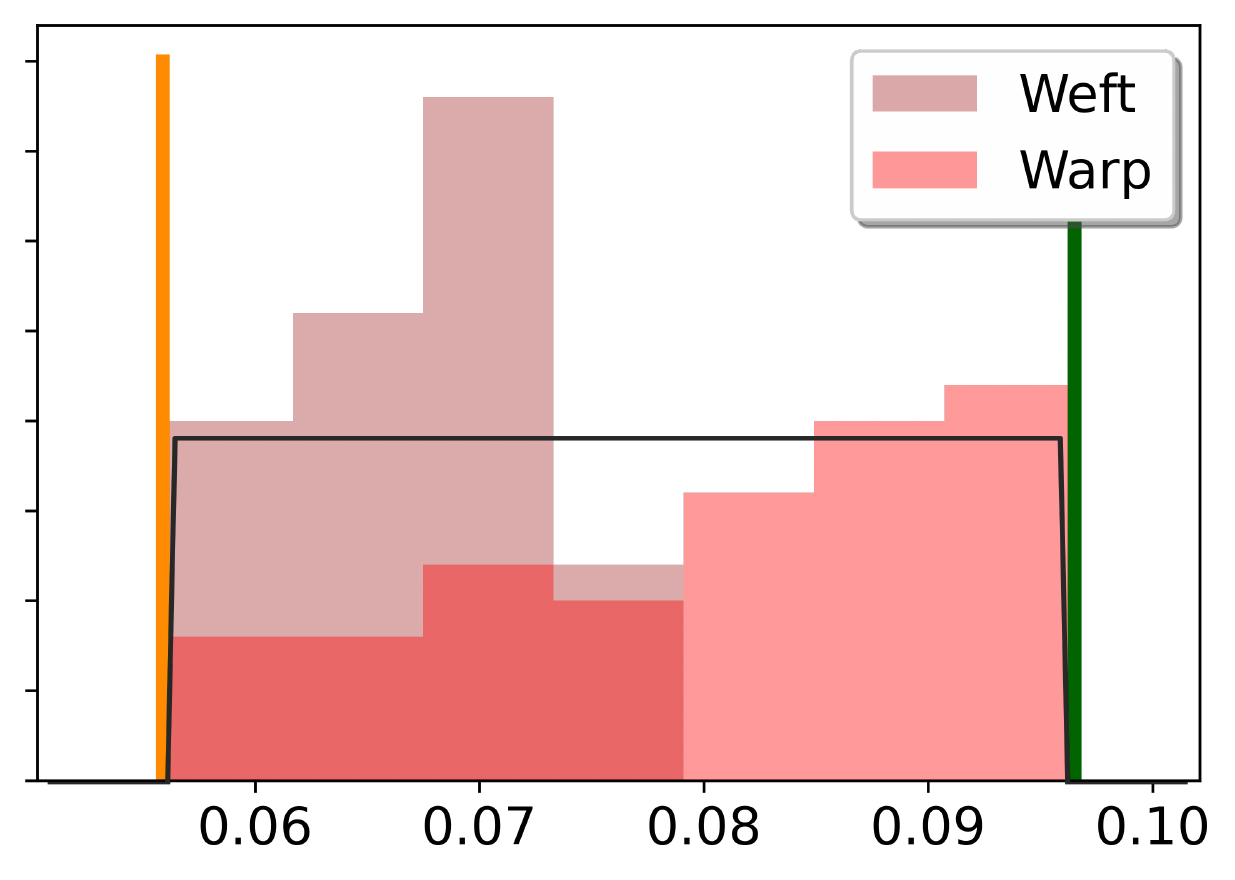}
            \caption{Tow width [\unit{\centi\meter}]}
            \label{fig:lhs_dist:w}
        \end{subfigure} 
        \begin{subfigure}{0.31\textwidth}
            \centering
            \includegraphics[width=\textwidth]{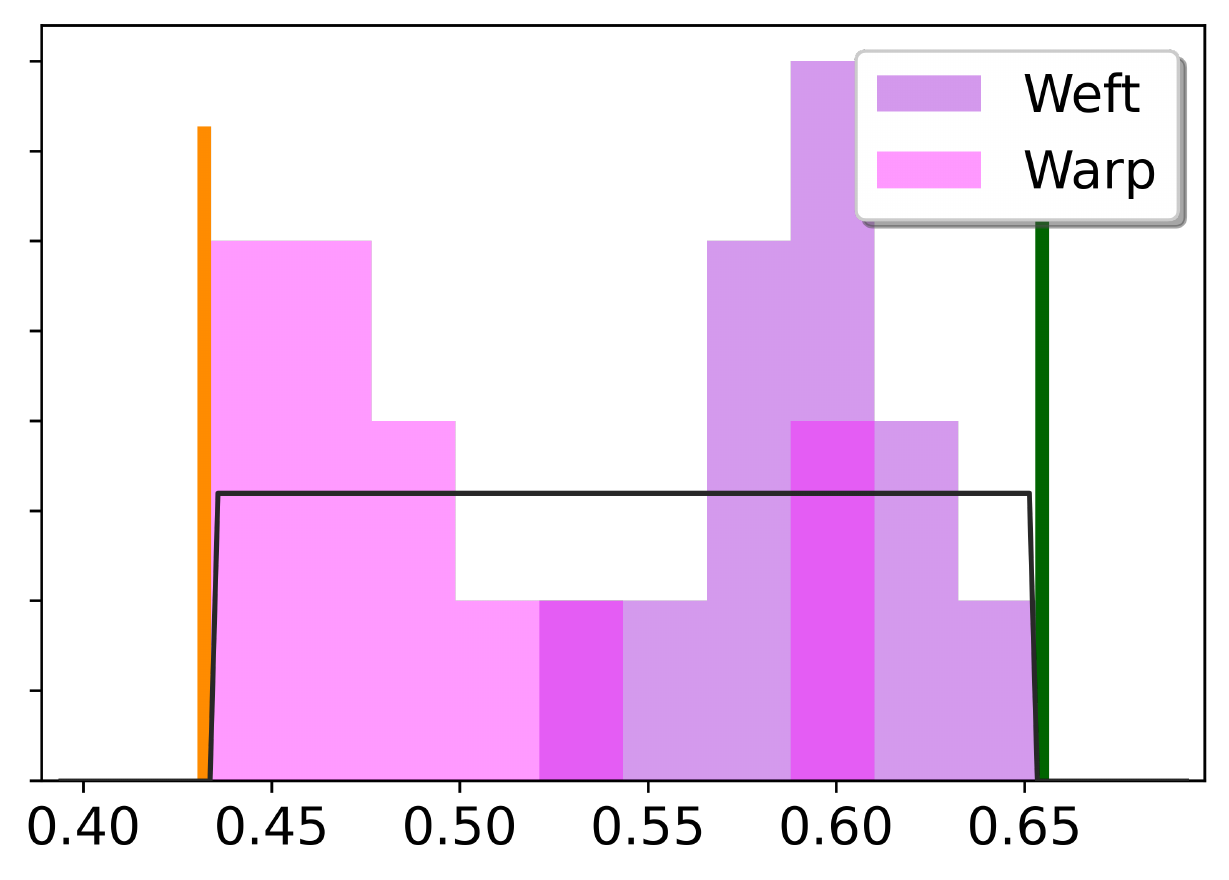}
            \caption{Undulation}
            \label{fig:lhs_dist:U}
        \end{subfigure}
        \caption[Composite geometry parameters harvested from XCT images.]{Composite geometric parameters harvested from XCT images. (a) XCT SL dataset sampled for: (b) tow thickness, (c) gap, (d) fiber packing, (e) width and (f) undulation. Measurements were collected separately for the warp and weft components to account for biasing and are depicted through separate histograms. The resulting distributions sampled through the LHS study are shown on each histogram, with bounds where appropriate.}
        \label{fig:lhs_dist}
    \end{figure}

    Next, we compare the properties generated from the parameterized analytical geometry to the properties from the SL XCT image. First, we extract geometric parameters throughout the SL image in warp- and weft-aligned domains, producing the sampled parameter distributions presented in \cref{fig:lhs_dist}. Thickness (\cref{fig:lhs_dist:t}) and fiber packing ratio (\cref{fig:lhs_dist:packing}) are well represented by normal distributions, capturing the sampling space with descriptors of their respective range, means, and standard deviations. The compact nature of the weave rendered many measurements of tow gap (\cref{fig:lhs_dist:gap}) at near indistinguishable zero length, prompting a log-normal descriptor to capture its bias maxima. Width (\cref{fig:lhs_dist:w}) and undulation (\cref{fig:lhs_dist:U}) parameters accentuate the geometric differences between the warp and weft orientations, as can be seen in the slightly bimodal distributions. Due to limitations in sampling description, we use a uniform distribution to capture both maxima of width and undulation. 

    \begin{figure*}
        \centering
        \begin{subfigure}[c]{0.69\textwidth}
            \centering
            \includegraphics[width=\textwidth]{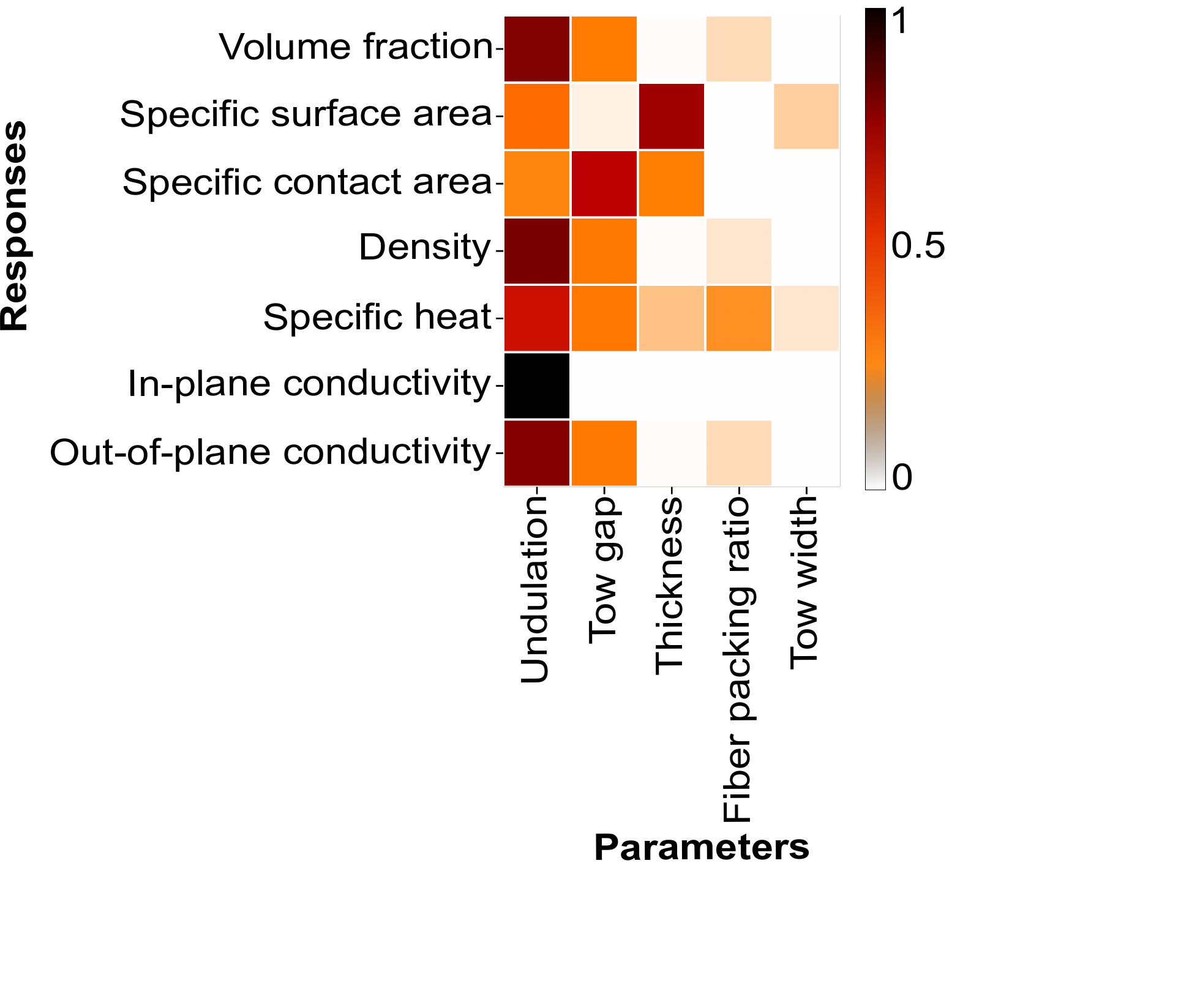}
            \caption{Sobol' indices}
            \label{fig:sobol:heatmap}
        \end{subfigure}
        \begin{subfigure}[t]{0.3\textwidth}
            \centering
            \includegraphics[width=\textwidth]{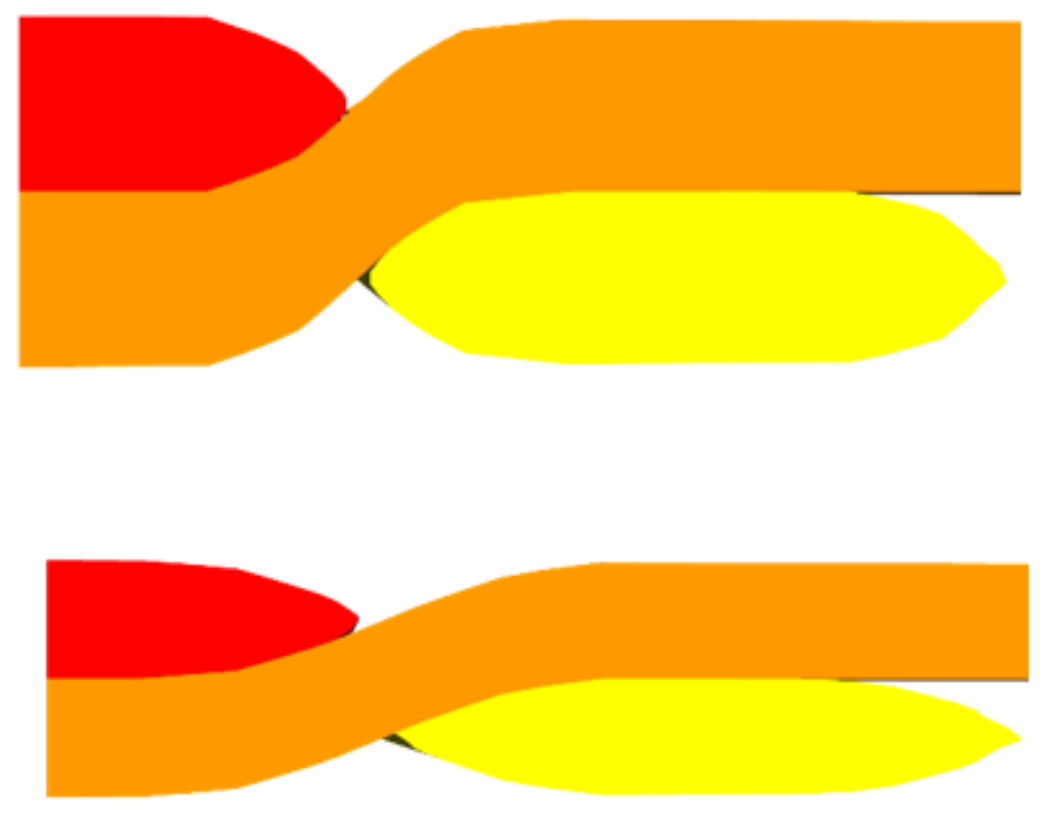}
            \caption{Undulation}
            \label{fig:sobol:undulation}
        \end{subfigure}
        \caption[Main effect Sobol' indices obtained using the PCE-developed surrogate model.]{(a) Main effect Sobol' indices obtained using the PCE-developed surrogate model. (b) Example simulated geometries generated using extremal values of the undulation parameter: the upper figure has a lower undulation (0.45) with a high degree of crimping, whereas the lower figure has a higher undulation value (0.65).}
        \label{fig:sobol}
    \end{figure*}

    We then performed a LHS sensitivity analysis of the analytical SL geometry, sampling from the parameter ranges described in \cref{fig:lhs_dist}. \Cref{fig:sobol:heatmap} shows the geometric parameters that dominate the effective composite properties through total Sobol' indices. Tow undulation and fiber packing ratio have the largest influences on the calculated thermal properties.

    Undulation has a large influence on OOP conductivity as directing more axially-conductive tows in the OOP direction would intuitively increase this conductivity. Shown in \cref{fig:sobol:undulation} is the geometrical variation from a maximum (0.65) and minimum (0.45) undulation parameter. One can see that undulation has a large bearing on the shape of the tows themselves, ultimately affecting thermal properties through fiber volume fraction and producing preferential pathways for thermal transport.
    
    Fiber packing ratio designates the packing of filaments within the tow, and as such, has clear influence on all properties dependent on the fibers: volume fraction, density, specific heat, IP and OOP conductivities. However, the sensitivity of IP conductivity on the packing ratio is overshadowed by the undulation. The tow gap describes the space between neighboring tows measured from the SL image geometry. Many of the tow gaps sampled are nearly zero in length, increasing the relative amount of heat conducted through the weave and allowing a more connected pathway for thermal transport. Again, for the IP direction, tows aligned with the thermal gradient dominate the thermal transport and perpendicular tows and the resin matrix, have relatively little importance. Tow thickness and width do not show significant influence on thermal properties through the Sobol' indices. Instead, they come through on properties more dependent on the resin such as specific heat and in morphology properties such as surface area. Weave volume fraction, and therefore fiber volume fraction, can be expressed in a closed-form and are independent of thickness and width under the translation method of creating the analytical geometry \cite{Collins2021}.

    \begin{figure*}
        \centering
        \begin{subfigure}{0.40\textwidth}
            \centering
            \includegraphics[width=\textwidth]{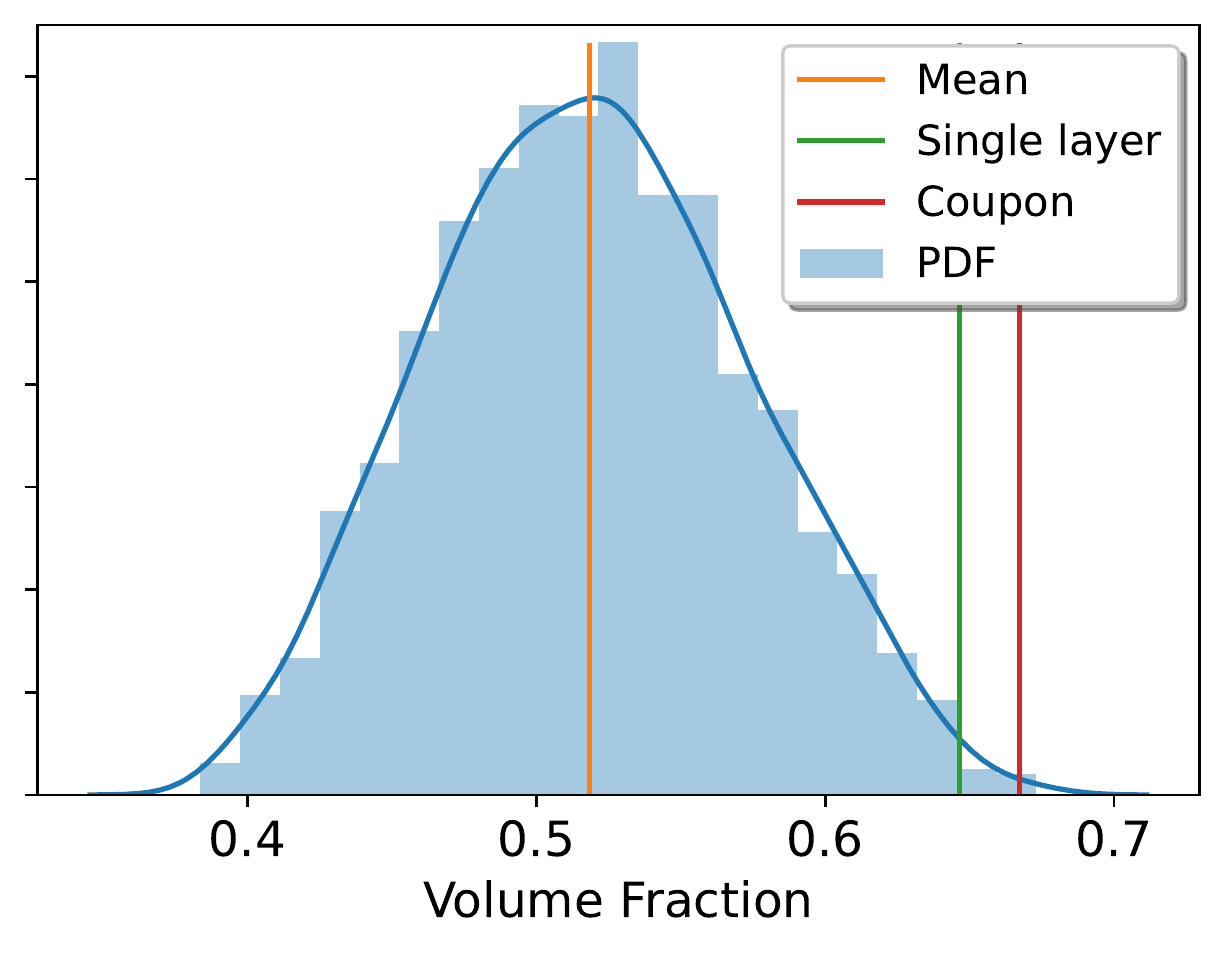}
            \caption{Fiber volume fraction}
            \label{fig:scatters:vf}
        \end{subfigure}
        \hspace*{.7cm}
        \begin{subfigure}{0.48\textwidth}
            \centering
            \includegraphics[width=\textwidth]{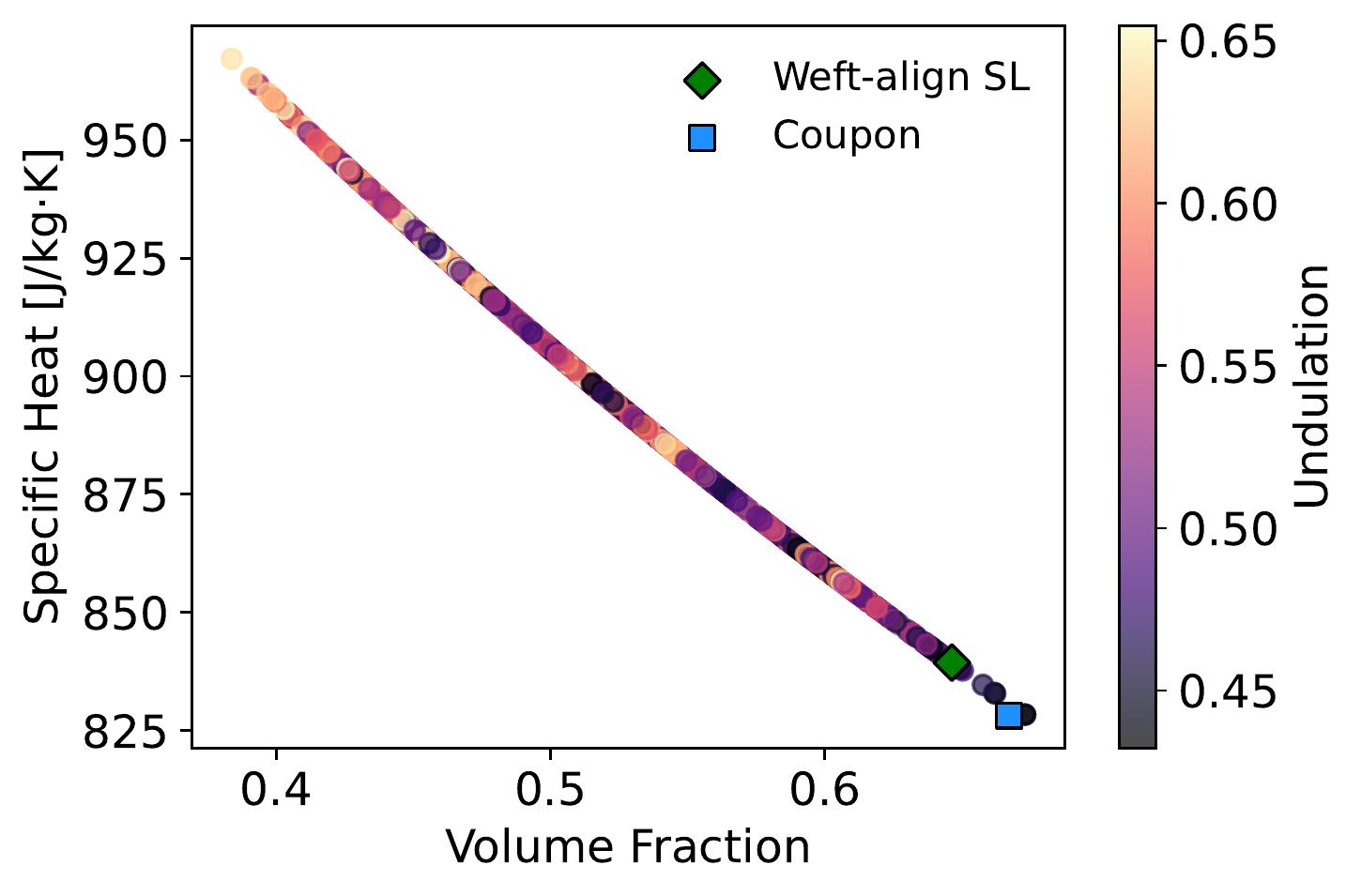}
            \caption{Specific heat}
            \label{fig:scatters:cp}
        \end{subfigure}	\\
        \begin{subfigure}{0.48\textwidth}
            \centering
            \includegraphics[width=\textwidth]{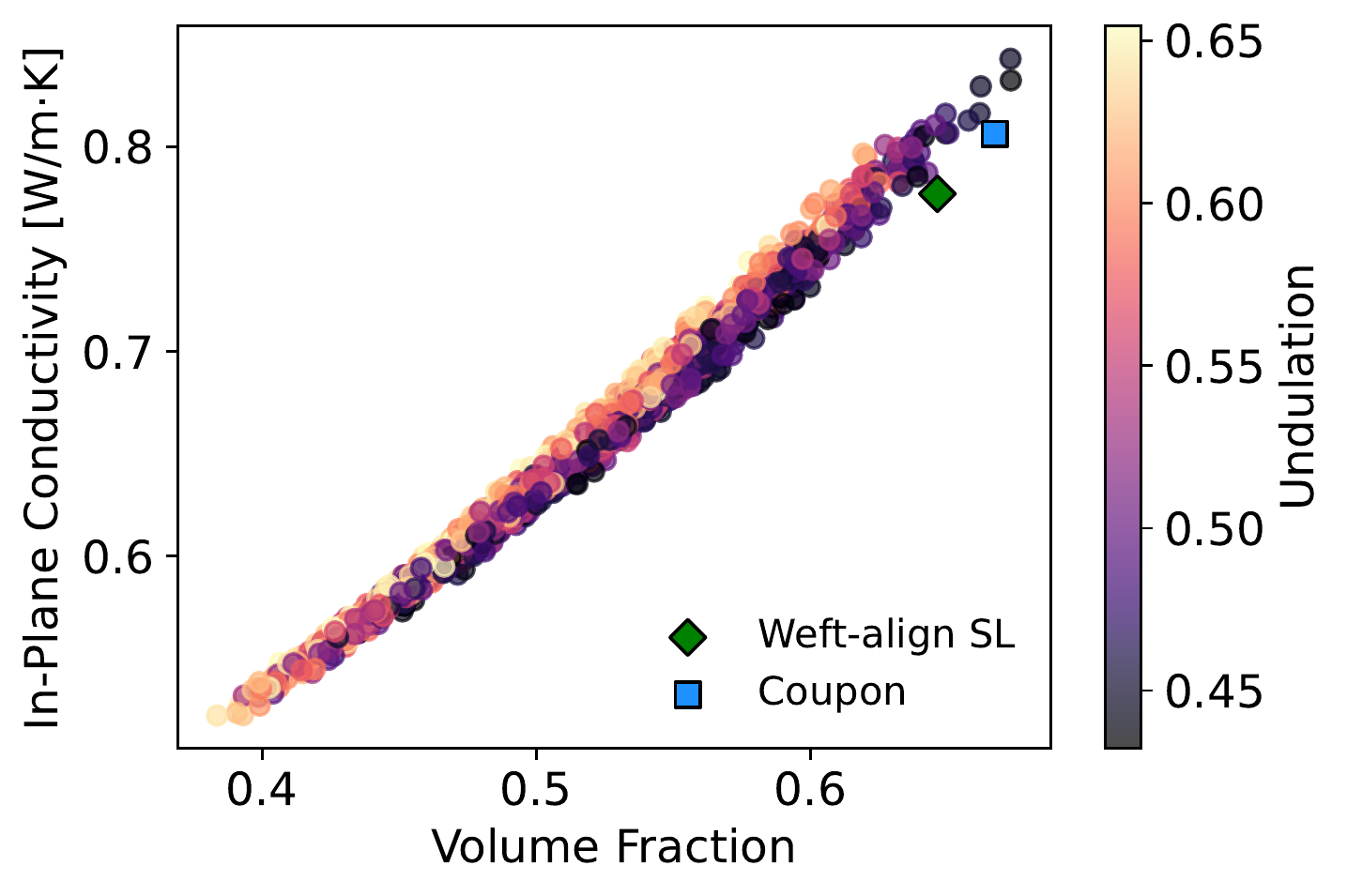}
            \caption{In-plane thermal conductivity}
            \label{fig:scatters:ip}
        \end{subfigure} 
        \begin{subfigure}{0.48\textwidth}
            \centering
            \includegraphics[width=\textwidth]{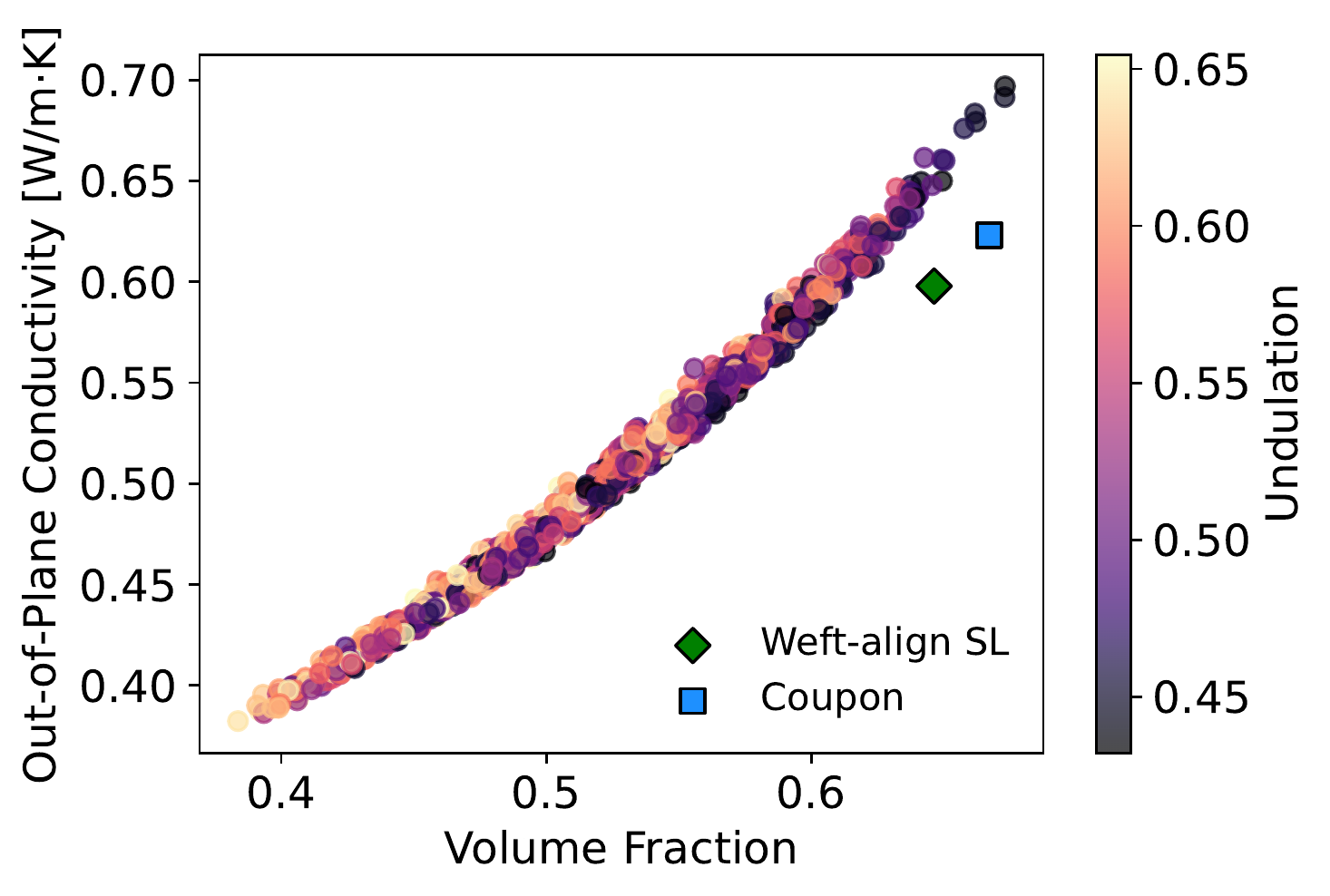}
            \caption{Out-of-plane thermal conductivity}
            \label{fig:scatters:oop}
        \end{subfigure} 
        \caption[Effective property results from the analytical geometry LHS study compared to image-based simulations.]{Effective property results from the analytical geometry LHS study compared to image-based simulations. (a) fiber volume fraction, (b) specific heat collected for all LHS realizations. (c) In-plane thermal conductivity, and (d) out-of-plane thermal conductivity. Scatter plots are colored by the undulation parameter to highlight secondary dependence.}
        \label{fig:scatters}
    \end{figure*}

    Depicted in \cref{fig:scatters} are the representative image-based SL and ML coupon results compared to those from the analytical study. \Cref{fig:scatters:vf} shows the distribution on fiber volume fractions. While the uncertainty sampling of the analytical geometry produces a normal distribution of volume fractions, the values extracted from both SL and ML coupons fall at the very high volume fraction tail of this distribution. This is because of the inability of our simple, single layer analytical geometry model to replicate the fabric layer nesting demonstrated by our very compact images. It is only when the highest fiber packing ratios, and lowest tow gap and undulation values are sampled that we can achieve volume fractions that approach that of the images. 

    Specific heat is simply a linear combination of the matrix and tow specific heats, weighted by the phase's mass fraction, shown in \cref{fig:scatters:cp}. A similar behavior to the volume fraction is seen, where the image-based simulation results fall on the outskirts of the analytical distribution. Undulation is the largest contributor to this spread, with the image-based values only occurring for small undulation values for the same reasons described previously for volume fraction. 

    Conductivity distributions are not as trivial as specific heat, with the curved and spread-out distributions shown in \cref{fig:scatters:ip,fig:scatters:oop}, as they cannot be exactly replicated as a linear combination of parameters.  The woven composites studied can be approximated fairly accurately with closed form models arising from theory \cite{Clayton1968}.

    IP conductivity can be closely approximated by a linear mixture model of resin and fiber, although the image-based data results in slightly lower conductivities than predicted by the analytical model (\cref{fig:scatters:ip}). Again, the image data is best aligned with the analytical model at high volume fractions and low undulation values.

    OOP conductivity (\cref{fig:scatters:oop}) derived from the image-based geometries show a much more significant deviation from the analytical model when compared to the IP values. This is because it is only through using a low undulation value that we can achieve the volume fractions observed in the XCT images. These undulation values are much lower than the average undulation of the images. Lower undulation means that the tows are more oriented in the OOP direction during their undulation, leading to a drastic increase in OOP conductivity (as tows are much more conductive along their axis than transversely). This results in the analytical model significantly overpredicting the image-based results. 

\section{Conclusion}

    We studied various approaches to calculating the effective thermophysical properties of woven composites using finite element simulations. First, we examined the mesoscale reconstruction of a X-ray computed tomography scan of a highly compacted multi-layer composite into a finite element simulation mesh. The resulting meshes were combined with material properties approximated using microscale effective media theory to compute effective properties of as-manufactured parts. Composite theory suggests that we should accurately estimate the bulk composite behavior though unit-cell simulations. Accordingly, we explored the dependence of effective properties on domain size.

    In this domain size study, we found that all subdomain sizes well represented the full coupon's properties in the in-plane direction in a mean sense, when enough subdomain samples are chosen, as long as tow orientation vectors are used.  However, any given single subdomain sample can deviate from the average behavior by as much as 25\%. While the variability of out-of-plane conductivity is much smaller than in-plane, the mean value is more biased away from the full coupon domain, by as much as 8\%. 
    
    The second portion of this study focused on the comparison of image-based simulation results with those calculated using analytical composite geometries. Parameters necessary to generate piecewise sinusoidal tow geometries were manually measured from the X-ray computed tomography scans and informed distributions sampled through an LHS study of 1500 realizations. A sensitivity analysis illustrated useful structure-property relations through Sobol' indices. Finally, a comparison between the analytical property distributions and the image-based results explain the applicability of analytical composite geometries in estimating bulk properties. 

    For the composite samples scanned and simulated in this study, the corresponding analytical geometries can represent the image-based geometry with select caveats. Due to the analytical geometry's inability to capture high compaction, layer nesting, and weave deformation seen in the manufactured samples, the synthetic geometries are not able to fully achieve comparable fiber volume fractions, even when using parameters measured from the X-ray computed tomography data. However, at low undulation values, where the fabric is highly crimped and the space between tows is reduced, the analytical geometries can obtain representative fiber volume fractions. With these combinations of parameters, the analytical geometries produce comparable results for thermal conductivity in the in-plane direction, as well as specific heat. Out-of-plane conductivity varies more at 10\% from the image-based, due to the very low undulation that comes with the analytical single-layer achieving a comparable volume fraction.  Through the comparisons performed, it is evident that the generated cross sections do not represent the compact nature of the real geometry. However, the simple model selected was chosen to produce valuable structure-property relations between tunable fabric characteristics and thermophysical performance.

    In analyzing the image-based results across two samples and a variety of domain sizes, we found 8\% maximum difference in the effective properties calculated. Performing simulations using a single-layer of a composite is a good approximation to the bulk properties considered. The material orientation modeled through the tow orientation vectors instead of coordinate-aligned, was not essential in this study, as the high weave volume fraction had much higher influence on thermal properties in the image-based geometries. The multi-layer coupon is reasonably represented by a sub unit-cell domain with consideration of high possible deviation in smaller domains. Higher resolution scans would allow for more reliable structure tensor calculations for local fiber orientation and therefore give a better idea on the role of material anisotropy for these samples. Future scans may focus on capturing microscale features without misrepresenting mesoscale effects on effective thermal property calculations. 

    The predictive models and techniques developed and studied herein have immense applicability to mesoscale studies of densely woven composites. The results presented show the value of simulating both image-based and analytical composite geometries for bulk property estimation. Limitations of this study suggest various topics for future studies. Studying more mesoscale features of composites, developing more realistic analytic geometries for parametric studies, and the application of next-generation machine learning methods for segmentation and material orientation are all areas of possible improvement. 

\section*{Declaration of competing interest}

    Financial support provided by Sandia National Laboratories. 

\section*{Acknowledgements}

    The authors gratefully acknowledge Bernadette Hernandez-Sanchez for help manufacturing the woven composites used in XCT imaging, which was performed by Christine C. Roberts.

    Supported by the Laboratory Directed Research and Development program at Sandia National Laboratories. This paper describes objective technical results and analysis. Any subjective views or opinions that might be expressed in the paper do not necessarily represent the views of the U.S. Department of Energy or the United States Government. This article has been authored by an employee of National Technology \& Engineering Solutions of Sandia, LLC under Contract No. DE-NA0003525 with the U.S. Department of Energy (DOE). The employee owns all right, title and interest in and to the article and is solely responsible for its contents. The United States Government retains and the publisher, by accepting the article for publication, acknowledges that the United States Government retains a non-exclusive, paid-up, irrevocable, world-wide license to publish or reproduce the published form of this article or allow others to do so, for United States Government purposes. The DOE will provide public access to these results of federally sponsored research in accordance with the DOE Public Access Plan https://www.energy.gov/downloads/doe-public-access-plan.

    This work was supported in part by the Department of Education through the Graduate Assistance in Areas of National Need Fellowship Program award P200A180050-19 at the University of Illinois Urbana-Champaign Department of Aerospace Engineering.

\printbibliography

\end{document}